\RequirePackage{ifpdf}
\ifpdf 
\documentclass[pdftex]{sigma}
\else
\documentclass{sigma}
\fi

\begin{document}

\allowdisplaybreaks

\renewcommand{\PaperNumber}{009}

\FirstPageHeading

\renewcommand{\thefootnote}{$\star$}

\ShortArticleName{A Unif\/ied Model of Phantom Energy and Dark Matter}

\ArticleName{A Unif\/ied Model of Phantom Energy\\ and Dark Matter\footnote{This
paper is a contribution to the Proceedings of the Seventh
International Conference ``Symmetry in Nonlinear Mathematical
Physics'' (June 24--30, 2007, Kyiv, Ukraine). The full collection
is available at
\href{http://www.emis.de/journals/SIGMA/symmetry2007.html}{http://www.emis.de/journals/SIGMA/symmetry2007.html}}}

\Author{Max CHAVES~$^\dag$ and Douglas SINGLETON~$^\ddag$}

\AuthorNameForHeading{M. Chaves and D. Singleton}

\Address{$^\dag$~Escuela de Fisica Universidad de Costa Rica, San Jose, Costa Rica}
\EmailD{\href{mailto:mchaves@cariari.ucr.ac.cr}{mchaves@cariari.ucr.ac.cr}}

\Address{$^\ddag$~Physics Department, CSU Fresno, Fresno, CA 93740-8031, USA}
\EmailD{\href{mailto:dougs@csufresno.edu}{dougs@csufresno.edu}}

\ArticleDates{Received November 01, 2007, in f\/inal form January
22, 2008; Published online January 30, 2008}

\Abstract{To explain the {\it acceleration} of the cosmological expansion
researchers have considered an unusual form of mass-energy
generically called dark energy. Dark energy has a~ratio of
pressure over mass density which obeys
$w = p / \rho < -1/3$. This form of mass-energy leads to
accelerated expansion. An extreme form of dark energy, called
phantom energy, has been proposed which has $w = p / \rho < -1$.
This possibility is favored by the observational data. The simplest
model for phantom energy involves the introduction of
a~scalar f\/ield with a negative kinetic energy term. Here we
show that theories based on graded Lie algebras naturally have such
a negative kinetic energy and thus give a model for phantom energy in a
less {\it ad hoc} manner. We f\/ind that the model
also contains ordinary scalar f\/ields and anti-commuting (Grassmann) vector
f\/ields which act as a form of two component dark matter. Thus from a
gauge theory based on a graded algebra we naturally obtained
both phantom energy and dark matter.}


\Classification{81R10; 81T10}

\section{Introduction}

Graded Lie algebras or Lie superalgebras (i.e.\ algebras having commuting
and anti-commuting generators) were at one time considered as models for a
more complete unif\/ied electroweak theory \cite{dondi} as well as Grand Unif\/ied Theories
\cite{taylor}. Such graded algebras had many attractive features:
vector and scalar bosons were contained within the same theory, the Weinberg angle was f\/ixed;
in some formulations the mass of the Higgs was f\/ixed. However these graded algebras generically
\cite{eccle} gave rise to negative kinetic energy terms
for some of the gauge f\/ields when the graded trace or supertrace was used.

Here we show that this negative kinetic energy
of the original graded algebras can be used to construct a
model for phantom energy \cite{caldwell,carroll}. In addition to the phantom f\/ield
there are other f\/ields which come from this model which act as dark matter. The
advantage of the combined phantom energy/dark matter model presented here is that
it is derived from a modif\/ied gauge principle (i.e.\ the gauge principle applied
to graded algebras) rather than being introduced by hand. This feature
f\/ixes the parameters, such as the coupling between the phantom energy and dark
matter, that are free in more phenomenological models.

Phantom energy is a form of dark energy which has a ratio of pressure over density
given by $w = p / \rho < -1$.   Dark energy in general is a cosmological
``f\/luid'' with $w < -1/3$, which gives rise to an accelerated
cosmological expansion. Dark energy was proposed to explain the accelerated
expansion observed in studies of distant type Ia supernova~\cite{riess,perlmutter}. There are various proposals as to the nature of dark energy:
a small, positive cosmological constant, quintessence~\cite{zlatev},
brane world models \cite{ddg,deff}, Chaplyin gas \cite{kam}, $k$-essense \cite{gonzalez1},
axionic tensor f\/ields \cite{gonzalez2} and others. A~good review can be
found in \cite{sahni}. Phantom energy is an extreme form of dark energy.  The simplest
model for phantom energy involves a scalar f\/ield with a negative kinetic energy term~\cite{caldwell}
\begin{gather}
\label{scalar-pe}
{\mathcal L}_p = -\frac{1}{2} (\partial _\mu \phi) (\partial ^\mu \phi) - V(\phi).
\end{gather}
The negative sign in front of the kinetic energy term makes this an
unusual f\/ield theory. Theories with negative kinetic energies have been
investigated theoretically starting with~\cite{bronn}. Other papers considering scalar f\/ields with
negative kinetic energies can be found in~\cite{nega-ke}. The main objection to
these negative kinetic energy theories is that quantum mechanically they
violate either conservation of probability or they have no
stable vacuum state due to an unbounded, negative energy density.
Although such unusual f\/ield theories are not ruled out~\cite{caldwell}, one can place constraints on them~\cite{cline}.
Despite the theoretical problems of a scalar f\/ield with a negative kinetic energy
term the reason to consider such a strange f\/ield theory is that recent
observations give  $-1.48 < w < -0.72$~\cite{hann} and thus favor $w < -1$. A recent
comparison of data from various sources can be found in~\cite{jass}.

The result $w<-1$ coming from the Lagrangian in \eqref{scalar-pe} depends not only on the
negative kinetic energy term, but also requires that the potential, $V(\phi )$, be present
and satisfy some conditions. One can calculate $p$ and $\rho$ from \eqref{scalar-pe}.
Assuming that the scalar f\/ield is spatially homogeneous enough so that only the
time variation is important one f\/inds
\begin{gather}
\label{w}
w = \frac{p}{\rho} = \frac{-\frac{1}{2}{\dot \phi} ^2 - V(\phi )}{-\frac{1}{2}{\dot \phi} ^2 + V(\phi )}.
\end{gather}
In order to have $w <-1$ the potential must satisfy $ \sqrt { 2 V (\phi ) } > | {\dot \phi} | \ge 0$.
We will show that it is possible, using graded algebras, to construct a f\/ield theory that satisf\/ies
these conditions and so gives rises to phantom energy. Unlike other models, the
negative kinetic term comes from the structure of the graded algebras rather than being put
in by hand. In addition there are other f\/ields which play the role of dark matter.

\section[Review of $SU(2/1)$ algebra]{Review of $\boldsymbol{SU(2/1)}$ algebra}

Here we brief\/ly review the graded algebra $SU(2/1)$.
The basic idea of using graded algebras to give phantom energy
works for larger graded algebras like $SU(N/1)$ with $N > 2$. We have taken $SU(2/1)$ for
simplicity.

We use the representation for $SU(2/1)$ which consists of the following eight
$3 \times 3$ matrices
\begin{alignat*}{4}
& \mbox{even:}\quad &&   T_{1}    =\frac{1}{2}\left(
\begin{array}
[c]{ccc}%
0 & 1 & 0\\
1 & 0 & 0\\
0 & 0 & 0
\end{array}
\right)  ,\qquad && T_{2}=\frac{1}{2}\left(
\begin{array}
[c]{ccc}%
0 & -i & 0\\
i & 0 & 0\\
0 & 0 & 0
\end{array}
\right) , &
\\
&&& T_{3}=\frac{1}{2}\left(
\begin{array}
[c]{ccc}%
1 & 0 & 0\\
0 & -1 & 0\\
0 & 0 & 0
\end{array}
\right), \qquad && T_{8}=\frac{1}{2}\left(
\begin{array}
[c]{ccc}%
1 & 0 & 0\\
0 & 1 & 0\\
0 & 0 & 2
\end{array}
\right), &
\\
& \mbox{odd:}\quad &&
 T_{4}    =\frac{1}{2}\left(
\begin{array}
[c]{ccc}%
0 & 0 & 1\\
0 & 0 & 0\\
1 & 0 & 0
\end{array}
\right)  ,\qquad && T_{5}=\frac{1}{2}\left(
\begin{array}
[c]{ccc}%
0 & 0 & -i\\
0 & 0 & 0\\
i & 0 & 0
\end{array}
\right)  ,&
\\
&&& T_{6}=\frac{1}{2}\left(
\begin{array}
[c]{ccc}%
0 & 0 & 0\\
0 & 0 & 1\\
0 & 1 & 0
\end{array}
\right) , \qquad && T_{7} =\frac{1}{2}\left(
\begin{array}
[c]{ccc}%
0 & 0 & 0\\
0 & 0 & -i\\
0 & i & 0
\end{array}
\right)  . &
\end{alignat*}
Except for $T_8$ this is the standard, fundamental
representation of $SU(3)$. The matrices on the f\/irst line above (i.e.\
$T_1$, $T_2$, $T_3$, $T_8$) are the even generators, and those on the second line
(i.e.\ $T_4$, $T_5$, $T_6$, $T_7$) are odd generators. The even generators satisfy
commutation relationships among themselves which can be written symbolically as
$[\mbox{even}, \mbox{even}] = \mbox{even}$. Mixtures of even and odd generators satisfy commutators of
the form $[\mbox{even}, \mbox{odd}] = \mbox{odd}$. Finally the odd generators satisfy anti-commutation
relationships of the form $\{ \mbox{odd}, \mbox{odd} \} = \mbox{even}$. The further details of the $SU(2/1)$
graded algebra can be found in the paper by Dondi and Jarvis~\cite{dondi} or
in Ecclestone~\cite{eccle}. The odd generators above are dif\/ferent than those
usually taken in the literature. The connection of the odd generators above with
those in \cite{dondi} is given by ${\bar Q} ^1 , Q_1 = T_4 \pm i T_5$
and ${\bar Q} ^2 , Q_2 = T_6 \pm i T_7$. In the rest of the article we will use the
convention that generators with indices from the middle of the alphabet ($i, j, k$)
are the even generators, $T_1$, $T_2$, $T_3$, $T_8$, while indices from the beginning of the
alphabet ($a, b, c$) are the odd generators $T_4$, $T_5$, $T_6$, $T_7$.

For the graded algebra one replaces the concept of the trace by the supertrace.
For $SU(2/1)$ this means that one writes some general element of the group as
\begin{align*}
\quad M =\left(
\begin{array}
[c]{cc}%
A_{2 \times 2} & B_{2 \times 1} \\
C_{1 \times 2} & d_{1 \times 1} \\
\end{array}
\right)  .
\end{align*}
The subscripts indicate the size of the sub-matrix. The supertrace is now def\/ined as
\begin{gather}
\label{supertrace}
{\operatorname {str}}(M) = {\operatorname {tr}} [A] - {\operatorname {tr}}[d]
\end{gather}
which dif\/fers from the regular trace due to the minus sign in front of $d$.

Later we will need the supertraces of the various products of the
eight generators $(T_i , T_a)$. We collect these results here. For products of
even generators we have
\begin{gather}
\label{even}
{\operatorname {str}}(T_i T_j) = \delta _{ij} \frac{1}{2} \qquad \text{except} \qquad
{\operatorname {str}}(T_8 T_8) = -\frac{1}{2}
\end{gather}
for the odd generators we have
\begin{gather}
\label{odd}
{\operatorname {str}} (T_4 T_5) = -{\operatorname {str}}(T_5 T_4) = \frac{i}{2} , \qquad
{\operatorname {str}} (T_6 T_7) = -{\operatorname {str}}(T_7 T_6) = \frac{i}{2}.
\end{gather}
All other supertraces of the product of two matrices are zero.

\section[Phantom energy and dark matter from an $SU(2/1)$ graded algebra]{Phantom energy and dark matter\\ from an $\boldsymbol{SU(2/1)}$ graded algebra}

In \cite{dondi} vector f\/ields were associated with the even
generators and scalar f\/ields with the odd generators as
\begin{gather}
\label{graded-alg}
A_\mu = i g A_\mu ^i T^{\rm even} _i, \qquad \phi  = -g \varphi ^a  T^{\rm odd} _a.
\end{gather}
The f\/ields $A_\mu ^i$ are regular commuting f\/ields while $\varphi ^a$ are Grassmann f\/ields.
In block form one can write \eqref{graded-alg} as
\begin{align*}
\quad A_M =\left(
\begin{array}
[c]{ccc}%
A_\mu ^3 + A_\mu ^8 & A_\mu ^1 - i A _\mu ^2 & \varphi ^4 - i \varphi ^5\\
A_\mu ^1 + i A_\mu ^2 & -A_\mu ^3 + A_\mu ^8 & \varphi ^6 - i \varphi^7\\
\varphi ^4 + i \varphi ^5 & \varphi ^6 + i \varphi ^7 & 2 A_\mu ^8
\end{array}
\right)  .
\end{align*}
In this fashion, {\it and by using the regular trace},  Dondi and
Jarvis \cite{dondi} showed that the Lagrangian
\begin{gather}
\label{g-lag}
{\mathcal L} = \frac{1}{2 g^2} {\operatorname {tr}} ( F_{MN} F^{MN} ) , \qquad F_{MN} = \partial_M A_N
-\partial _N A_M + [A_M, A_N] ,
\end{gather}
reduced to an $SU(2) \times U(1)$ Yang--Mills Lagrangian for $A_\mu$ and a
Higgs-like Lagrangian for $\phi$. In \eqref{g-lag} we use a dif\/ferent overall
sign for the Lagrangian as compared to \cite{dondi}. This comes because we
have chosen dif\/ferent factors of $i$ in the vector potentials def\/ined
below in \eqref{graded-alg2}. Using such an $SU(2/1)$ algebra gave
a more unif\/ied electroweak theory. The theory based on the graded $SU(2/1)$ algebra
was more unif\/ied in two ways: (i) There was only one coupling constant $g$ rather than
two separate coupling constant in the usual Standard Model based on $SU(2) \times U(1)$.
Thus in the Standard Model based on the graded $SU(2/1)$ algebra the Weinberg angle
was predicted rather than being an input parameter. (ii) Second the theory based on
the graded $SU(2/1)$ algebra automatically had a scalar f\/ield coming from the odd
terms in \eqref{graded-alg}.

However on further investigation \cite{eccle} there were problems with
using of the graded $SU(2/1)$ algebra to construct an electroweak
theory. If in \eqref{g-lag} one used the correct
$SU(2/1)$ invariant supertrace then the Yang--Mills part of the reduced Lagrangian
would have the wrong sign for the kinetic term for the $U(1)$ gauge f\/ield
and the kinetic energy term for the scalar f\/ield
would be lost.

Here we use these apparent negative features
to construct a model for phantom energy. Instead of making the association between
even/odd generators and vector/scalar f\/ields made in \eqref{graded-alg} we take
the opposite choice \cite{max}
\begin{gather}
\label{graded-alg2}
A_\mu = i g A_\mu ^a T^{\rm odd} _a  , \qquad \phi = - g \varphi  ^i T^{\rm even} _i.
\end{gather}
Because of the reversal of roles relative to \eqref{graded-alg} the f\/ields $A_\mu ^a$
are Grassmann f\/ields while $\varphi ^i$ are regular, commuting f\/ields. Then taking the
correct, $SU(2/1)$ invariant, supertrace we f\/ind that one of the scalar f\/ields develops
a negative kinetic energy term in addition to having a potential term which is positive
def\/inite. Thus the graded algebra gives rise to a phantom f\/ield.

With the choice in \eqref{graded-alg2} the Lagrangian in \eqref{g-lag} reduces as
follows \cite{max}
\begin{gather}
\label{g-lag2}
{\mathcal L} = \frac{1}{2 g^2} {\operatorname{str}} ( F_{MN} F^{MN} ) =
\frac{1}{2g^{2}}{\operatorname{str}}\left[  \left(  \partial_{\lbrack\mu}A_{\nu]}+[A_{\mu},A_{\nu}]\right)
^{2}\right] +\frac{1}{g^{2}}{\operatorname{str}}\left[  \left(  \partial
_{\mu}\phi  + [A_{\mu},\phi ]\right)  ^{2}\right].
\end{gather}
We have introduced the notation
$\partial_{\lbrack\mu}A_{\nu]} = \partial _\mu A_\nu - \partial _\nu A_\mu$.  Note
that in comparison to other works such as \cite{dondi} and \cite{eccle} we have not
introduced extra Grassmann coordinates, $\zeta^\alpha$ in addition to the normal
Minkowski coordinates $x^\mu$. Thus in~\cite{dondi} and~\cite{eccle} coordinates and
indices ran over six values~-- four Minkowski and two Grassmann. The f\/inal result in~\eqref{g-lag2}
can be obtained from~\cite{dondi} by dropping the Grassmann coordinates.

We f\/irst focus  on the scalar term in \eqref{g-lag2}. Inserting $\phi$ and $A_\mu$ from
\eqref{graded-alg2} into the last term in~\eqref{g-lag2} we f\/ind~\cite{max}
\begin{gather}
{\mathcal L}_S =  \frac{1}{g^{2}}{\operatorname{str}}\left[
\left(  \partial_{\mu}\phi+[A_{\mu},\phi]\right)  ^{2}\right] \nonumber \\
\phantom{{\mathcal L}_S}{} = {\operatorname{str}}\left[  \left(  \partial_{\mu}\varphi
^{8}T_{8}+igA_{\mu}^{a}\varphi^{8}[T_{a},T_{8}]\right)  ^{2}\right]
+ {\operatorname{str}}\left[  \left(  \partial_{\mu}\varphi
^i T_i +igA_{\mu}^{a}\varphi^i[T_{a},T_i]\right)  ^{2}\right].\label{scalar}
\end{gather}
The f\/irst term in \eqref{scalar} takes the form of a phantom energy f\/ield.
Expanding the f\/irst term in~\eqref{scalar} gives
\begin{gather}
\label{phantom}
\mathcal{L}_{\rm Phantom} =  {\operatorname{str}}\left[  \left(  \partial_{\mu}\varphi
^{8}T_{8}-gA_{\mu}^{4}\varphi^{8}T_{5}/2+gA_{\mu}^{5}\varphi^{8}%
T_{4}/2 -gA_{\mu}^{6}\varphi^{8}T_{7}/2 + gA_{\mu}^{7}\varphi^{8}%
T_{6}/2\right)  ^{2}\right].\!\!
\end{gather}
We have used the representation of the $SU(2/1)$ matrices from the previous
section to evaluate the commutators. Using the supertrace results from \eqref{even}
and \eqref{odd} the expression in \eqref{phantom} yields
\begin{gather}
\label{phantom2}
\mathcal{L}_{\rm Phantom}   =  -\frac{1}{2}(\partial_{\mu}\varphi^{8})^{2}- \frac{1}{16}g^{2}(\varphi
^{8})^{2}\left( A_{\mu}^{+}A^{- \mu} - A_{\mu}^{-}A^{+ \mu} + B_{\mu}^{+}B^{- \mu}
- B_{\mu}^{-}B^{+ \mu} \right)
\end{gather}
with $A_{\mu}^\pm=A_{\mu}^{4} \pm iA_{\mu}^{5}$
and $B_{\mu}^\pm =A_{\mu}^{6} \pm iA_{\mu}^{7}$. Both $A_{\mu} ^\pm$
and $B_{\mu} ^\pm$ are Grassmann so the last line in~\eqref{phantom2}
can be written
\begin{gather}
\label{phantom3}
\mathcal{L}_{\rm Phantom} = -\frac{1}{2}(\partial_{\mu}\varphi^{8})^{2}- \frac{1}{8}g^{2}(\varphi
^{8})^{2}\left( A_{\mu}^{+}A^{- \mu} + B_{\mu}^{+}B^{- \mu} \right).
\end{gather}
This is of the form of the phantom energy Lagrangian in \eqref{scalar-pe} but with the
potential involving not only the scalar f\/ield, $\varphi ^8$, but Grassmann vector f\/ields,
$A_\mu ^\pm$ and $B_\mu ^\pm$. We will discuss these shortly. The minus sign in front of the kinetic
energy term comes from taking the $SU(2/1)$ invariant supertrace rather
than the ordinary trace (see the second supertrace result in \eqref{even}).

We next focus on the other scalar f\/ields, $\varphi ^i$, $i =1,2,3$ which come from the
second term in~\eqref{scalar}. The calculation proceeds as in equations \eqref{phantom},
\eqref{phantom2} but with $\varphi ^8$ replaced by $\varphi ^i$, $i= 1, 2, 3$. For example
for $\varphi ^1$ \eqref{phantom} becomes
\begin{gather}
\label{dm}
\mathcal{L}_{\varphi ^1}  = {\operatorname{str}}\left[  \left(  \partial_{\mu}\varphi
^{1}T_{1}+gA_{\mu}^{4}\varphi^{1}T_{7}/2-gA_{\mu}^{5}\varphi^{1}%
T_{6}/2 -gA_{\mu}^{6}\varphi^{1}T_{5}/2 + gA_{\mu}^{7}\varphi^{1}%
T_{4}/2\right)  ^{2}\right]
\end{gather}
and \eqref{phantom2} becomes
\begin{gather}
\label{dm2}
\mathcal{L}_{\varphi^1} =  \frac{1}{2}(\partial_{\mu}\varphi^{1})^{2}- \frac{1}{8}g^{2}(\varphi
^{1})^{2}\left( A_{\mu}^{+}A^{- \mu} + B_{\mu}^{+}B^{- \mu} \right).
\end{gather}
There are two keys points: the kinetic term for $\varphi ^1$ is positive since
${\operatorname{str}} (T_1 T_1) = + 1/2$, and the potential term is the same as
for $\varphi ^8$. The other two even scalar f\/ields follow
the same pattern so that in total one can write
\begin{gather}
\label{dm3}
\mathcal{L}_{DM} = \frac{1}{2}(\partial_{\mu}\varphi^{i})^{2}- \frac{1}{8}g^{2}(\varphi
^{i})^{2}\left( A_{\mu}^{+}A^{- \mu} + B_{\mu}^{+}B^{- \mu} \right),
\end{gather}
where $i$ is summed from 1 to 3. Thus the total scalar f\/ield Lagrangian resulting from
\eqref{scalar} is the sum of~\eqref{phantom3} and~\eqref{dm3}. The scalar f\/ield in
\eqref{phantom3} has the ``wrong'' sign for the kinetic term and acts as a phantom
f\/ield. The scalar f\/ields in~\eqref{dm3} are ordinary scalar f\/ield which we will interpret
as a dark matter candidate. The phantom f\/ield and dark matter f\/ields are coupled
through the $A^\pm _\mu$ and $B^\pm _\mu$ f\/ields. Thus our model provides a coupling
between phantom energy and dark matter. Other models have been considered \cite{cai} where
there is coupling between dark/phantom energy and dark matter.

We will now examine the Grassmann vector f\/ields, $A^4 _\mu$,
$A^5 _\mu$, $A^6 _\mu$, $A^7 _\mu$. The f\/inal Lagrangian for these f\/ields will
have a nonlinear interaction between the $A^{\pm}_\mu$ and $B^{\pm}_\mu$ f\/ields.
In analogy with QCD we argue that these f\/ields form permanently conf\/ined condensates like
$\langle A^4 _\mu A^5 _\mu \rangle$ or~$\langle A^+ _\mu A^- _\mu \rangle$.
These then supply potential (mass-like)
terms for the phantom energy and scalar f\/ields of \eqref{phantom3} and \eqref{dm3}.
This also avoids violation of the spin-statistics theorem since these condensates
have bosonic statistics (they are composed of two Grassmann f\/ields) and integer
spin (they are composed of two integer spin f\/ields).
Having a potential term is crucial for the interpretation of $\varphi ^8$ as a
phantom energy f\/ield, since for a massless, non-interacting scalar f\/ield reversing the
sign of the kinetic energy term does not lead a phantom f\/ield with $w<-1$ as can be seen
from \eqref{w} if $V( \phi) =0$.
From \eqref{g-lag2} the vector part of the Lagrangian can be expanded as
\begin{gather}
\label{vector}
\mathcal{L}_V =  - \frac{1}{2}{\operatorname{str}}
\left[  \left(  \partial_{\lbrack\mu}A^a _{\nu]} T_a \right) ^2 \right]
+ \frac{g^2}{2} {\operatorname{str}}\left[ \left( A^a _\mu A^b _\nu \{ T_a,T_b \} \right)
^{2}\right] = \mathcal{L}_{V1} + \mathcal{L}_{V2}.
\end{gather}
The commutator has become an anticommutator due to the Grassmann nature of the $A^a _\mu$'s. Also
note that there is no cubic cross term between the derivative and anticommutator part. This
comes about since the anticommutator, $\{ T_a , T_b \}$ results in even generators, and
the supertrace between odd and even generators vanishes. $\mathcal{L}_{V1}$ is
a kinetic term for the f\/ields and $\mathcal{L}_{V2}$ a~potential term. We will now consider each
of these in turn.

The kinetic part can be written explicitly as
\begin{gather}
\label{v-kinetic}
\mathcal{L}_{V1} =  - \frac{1}{2}\,{\operatorname{str}}
\left[  \left(  \partial_{\lbrack\mu}A^4 _{\nu]} T_4 + \partial_{\lbrack\mu}A^5 _{\nu]} T_5
+\partial_{\lbrack\mu}A^6 _{\nu]} T_6 + \partial_{\lbrack\mu}A^7 _{\nu]} T_7\right) ^2 \right].
\end{gather}
Due to the property of the supertrace of the odd generators given in \eqref{odd} it is
only the cross terms between $T_4$, $T_5$ and $T_6$, $T_7$ which survive.
\begin{gather}
\label{v-kinetic2}
\mathcal{L}_{V1} = - \frac{i}{2}
\left(  \partial_{\lbrack\mu}A^4 _{\nu]} \partial_{\lbrack\mu}A^5 _{\nu]}
+\partial_{\lbrack\mu}A^6 _{\nu]} \partial_{\lbrack\mu}A^7 _{\nu]} \right)
= - \frac{1}{4}
\left(  \partial_{\lbrack\mu}A^- _{\nu]} \partial_{\lbrack\mu}A^+ _{\nu]}
+\partial_{\lbrack\mu}B^- _{\nu]} \partial_{\lbrack\mu}B^+ _{\nu]} \right),
\end{gather}
where we have used the anticommutating properties of the $A ^a _\mu$'s.
In the last step we have replaced the $A ^a _\mu$ by $A^\pm _\mu$ and
$B^\pm _\mu$. This kinetic part is reminiscent of the kinetic terms for
a charged (i.e.\ complex) vector f\/ield.

Next we work out the form of the interaction terms coming from
${\mathcal L}_{V2}$. We do this explicitly for $A^4 _\mu$; the
results for the other vectors f\/ields can be obtained in a similar manner.
The $A^a _\mu = A^4 _\mu$ part of $\mathcal{L}_{V2}$ expands like
\begin{gather}
\label{vector-int}
{\mathcal L}_{V2} = \frac{g^2}{2} {\operatorname{str}}\left[ \big( A^4 _\mu A^4 _\nu \{ T_4,T_4 \}
+ A^4 _\mu A^5 _\nu \{ T_4,T_5 \} + A^4 _\mu A^6 _\nu \{ T_4,T_6 \}
+ A^4 _\mu A^7 _\nu \{ T_4,T_7 \}
\big)^{2}\right].
\end{gather}
Using the explicit representations of the odd matrices we have $\{T_4 , T_4 \} =
(T_3 + T_8)/2$, $\{ T_4 , T_5 \} = 0$, $\{ T_4 , T_6 \} = T_1 /2 $,
$\{ T_4 , T_7 \} = -T_2 / 2$. Squaring and using the supertrace results
of \eqref{even} one f\/inds that \eqref{vector-int} becomes
\begin{gather}
\label{vector-int2}
{\mathcal L}_{V2} = \frac{g^2}{16}\left( A^4 _\mu A^6 _\nu A^{4 \mu} A^{6 \nu}
+ A^4 _\mu A^7 _\nu A^{4 \mu} A^{7 \nu} \right).
\end{gather}
Note that there is no quartic term in $A^4 _\mu$ since the contributions from
$T_3$ and $T_8$ cancel. The contribution from $A^5 _\mu$ looks the same as
\eqref{vector-int2} but with $A^4 _\mu \rightarrow A^5 _\mu$. The $A^6 _\mu$
and $A^7 _\mu$ terms can be obtained by making the exchange $A^4 _\mu
\leftrightarrow A^6 _\mu$ and $A^5 _\mu \leftrightarrow A^7 _\mu$. Using the
Grassmann character of the $A^a _\mu$'s one can see that the $A^4 _\mu$ and $A^6 _\mu$
contributions, and also the $A^5 _\mu$ and $A^7 _\mu$ contributions are the same.
In total the interaction part of the vector Lagrangian can be written as
\begin{gather}
\label{vector-int3}
{\mathcal L}_{V2} = \frac{g^2}{16} \left( A^+ _\mu B^+ _\nu A^{- \mu} B^{- \nu} + A^+ _\mu B^- _\mu
A^{- \nu} B^{+ \nu} \right).
\end{gather}
In the last line we have written the interaction in terms of $A^{\pm} _\mu$, $B^{\pm} _\mu$.

The total Lagrangian for the vector Grassmann f\/ields is, ${\cal L}_{V1} + {\cal L}_{V2}$, where
${\cal L}_{V1}$ is a kinetic term and ${\cal L}_{V2}$ gives a nonlinear
interaction term between $A^\pm _\mu$ and $B^\pm _\mu$. We assume that the interaction
is strong enough that the f\/ields, $A^\pm _\mu$ and $B^\pm _\mu$ are permanently conf\/ined into
condensates
\begin{gather}
\label{vev}
\langle A^+ _\mu A^{- \mu} \rangle = \langle B^+ _\mu B^{- \mu} \rangle = v.
\end{gather}
From the symmetry between the $A^{\pm}_\mu$ and $B^{\pm} _\mu$ f\/ields we have
set their vacuum expectation value to be equal. This conjectured condensation
is similar to the {\it gauge variant}, mass dimension~2 condensate, in regular
Yang--Mills theory, $\langle {\cal A}^a _\mu {\cal A}^{a \mu} \rangle$. Despite being
{\it gauge variant} this quantity has been shown \cite{boucaud} to
have real physical consequences in
QCD. Here ${\cal A}^a _\mu$ is a normal $SU(N)$ Yang--Mills f\/ield. In \cite{kondo} a BRST-invariant
mass dimension~2 condensate was constructed which was a combination of the quadratic
gauge f\/ield term~-- $\langle {\cal A}^a _\mu {\cal A}^{a \mu} \rangle$~--
plus a quadratic Fadeev--Popov~\cite{fp} ghost f\/ield term~-- $i \alpha \langle {\cal C}^a  {\bar{\cal C}}^a \rangle$~--
where $\alpha$ was a gauge parameter. In the Landau gauge, $\alpha =0$, this reduced to a
pure quadratic gauge f\/ield condensate $\langle {\cal A}^a _\mu {\cal A}^{a \mu} \rangle$.
Note that the ghost f\/ields, ${\cal C}^a $, ${\bar{\cal C}}^a$, are bosonic,
Grassman f\/ields. This mass dimension~2 condensate gives the gluon a mass~\cite{ds}. Estimates have been made for
$\sqrt{\langle {\cal A}^a _\mu {\cal A}^{a \mu} \rangle}$ using lattice methods
\cite{boucaud,shakin}, analytical techniques \cite{dudal} or some mixture.
All these methods give a condensate value in the range
$\sqrt{\langle {\cal A}^a _\mu {\cal A}^{a \mu} \rangle} \approx 1$~GeV.
From the similarities between the regular gauge f\/ield condensate
of \cite{boucaud,kondo} and that on the left
hand side of \eqref{vev} we estimate the vacuum expectation value as
$v \approx 1$~GeV$^2$.

Inserting these vacuum expectation values into \eqref{phantom3} yields
\begin{gather}
\label{phantom4}
\mathcal{L}_{\rm Phantom} = -\frac{1}{2}(\partial_{\mu}\varphi^{8})^{2}- \frac{v}{4}g^{2}(\varphi
^{8})^{2}.
\end{gather}
This is of the form \eqref{scalar-pe} with $V(\varphi ^8) =  \frac{v}{4}g^{2}(\varphi
^{8})^{2}$. This will give phantom energy with $w < -1$ if
$\frac{g}{2} |\varphi ^8 | {\sqrt{2 v}} > | \dot \varphi ^8 |$.
If the vacuum expectation value,~$v$, changes over time it is possible
to cross into (out of) the phantom regime if $v$ increases (decreases).
Thus whether one has phantom energy or not would depend on the dynamical
evolution of~$v$. Such models, where one crosses the ``phantom divide'', have been
considered in \cite{feng1}. In such models it is usually the sign in front of the
kinetic energy term that is modif\/ied, whereas in the present case it is a modif\/ication
of the potential which causes the transition between phantom and non-phantom phases.
Further extensions of these ``quintom'' models can be found in~\cite{feng2}.

Inserting the vacuum expectation values into the Lagrangian for the scalar f\/ields
$\varphi _1$, $\varphi _2$, $\varphi _3$, equation \eqref{dm3} becomes
\begin{gather}
\label{dm4}
\mathcal{L}_{DM} = \frac{1}{2}(\partial_{\mu}\varphi^{i})^{2}- \frac{v}{4}g^{2}(\varphi
^{i})^{2}.
 \end{gather}
The Lagrangian for these f\/ields is for a standard, non-interacting scalar with mass
$m= \frac{g}{2}{\sqrt{2 v}}$. These massive scalar f\/ields could be
cold dark matter if $m$ (i.e.~$v$) is chosen appropriately. For example,
using the similarity between the condensate of \eqref{vev} and the mass dimension
condensate of \cite{boucaud,kondo} one might set $v \approx 1$~GeV$^2$.
This would given $m \approx 1$ GeV making $\varphi ^a$ a viable, cold dark matter candidate.

The original Lagrangian \eqref{g-lag2} has no coupling to the
usual Standard Model f\/ields except through gravity. This would explain
why these phantom energy and dark matter f\/ields have not been seen since they
could only be detected through their gravitational inf\/luence. However if this
is the path nature chooses it would be hard, if not impossible, to get any kind
of experimental signal of these phantom energy/dark matter candidates. One could introduce
some ef\/fective coupling between the phantom energy/dark matter f\/ields of \eqref{g-lag2}
and the usual Standard Model f\/ields. More rigorously one might try to use some larger
$SU(N/1)$ group, but with some of the vector f\/ields associated with the even generators
and some associated with the odd generators and similarly for the scalar f\/ields. In this way
it might be possible to have a new kind of ``Grand Unif\/ied Theory'': from a single Lagrangian
one could have Standard Model gauge f\/ields as well as new f\/ields that would be phantom energy and
dark matter candidates, instead of extra Grand Unif\/ied gauge bosons.

The Grassmann vector f\/ields are an odd feature of this model since they would violate the
spin-statistics theorem. These Grassmann vector f\/ields are similar to the Fadeev--Popov ghosts~\cite{fp}:
scalar f\/ields with Fermi--Dirac statistics. The Fadeev--Popov ghosts do not violate the spin-statistics
since they never appear as asymptotic states. In order to avoid having
the Grassmann vector f\/ields violate the spin-statistics theorem, we have postulated that
the composite states, $A^+ _\mu A^{- \mu}$ and $B^+ _\mu B^{- \mu}$ are permanently conf\/ined so that
the particles associated with $A^\pm _\mu$ and $B^\pm _\mu$ never appear as asymptotic states.
Since the composites are ordinary f\/ields (integer spin with bosonic statistics) violation
of the spin-statistics theorem is avoided. These vectors f\/ields act as a
second dark matter component in addition to the three scalar f\/ields~$\varphi _i$.
There have been other recent proposals for dark matter candidates with non-standard
relationships between spin and mass dimension. In~\cite{grumiller} a spin~1/2 dark matter
candidate was proposed which has mass dimension~1. In the present case our vector f\/ields,
$A^{\pm} _\mu$, $B^{\pm} _\mu$, have the same mass dimension (i.e.~1) and statistics
(fermionic) as the dark matter candidate in \cite{grumiller}, and only dif\/fer in the
value of spin~-- 1 versus~1/2.

\section{Conclusions}

\looseness=-1
We have given a model for phantom energy using a modif\/ication of the
graded Lie algebras models which attempted to give a more unif\/ied electroweak
theory, or Grand Unif\/ied theories. Despite interesting features of the original graded Lie
algebra models (e.g.~prediction of the Weinberg angle and having
both vectors and scalars coming from the same Lagrangian) they
had shortcomings. Chief among these was that if one used the correct $SU(N/1)$ invariant
supertrace then some of the vector f\/ields had the wrong sign for the kinetic energy term in
the Lagrangian. In the original models the vector f\/ields were associated with the even
generators of the algebra and the scalars f\/ields were associated with the odd
generators. Here we took the reverse identif\/ication (scalar f\/ield $\rightarrow$ even generators
and vector f\/ield $\rightarrow$ odd generators) which led to the wrong sign kinetic energy
term coming from a scalar f\/ield rather than from a vector f\/ield. The wrong sign scalar f\/ield, $\varphi ^8$,
gives a model of phantom energy, while the other scalar f\/ields, $\varphi ^i$, and the
vector f\/ields, $A^a _\mu$, act as dark matter components. In the way our model is
formulated here all the f\/ields are truly dark in that they have no coupling to any of
the Standard Model f\/ields and would thus only be detectable via their gravitational interaction.
This would make the experimental detection of these dark f\/ields impossible through
non-gravitational interactions. However the above is intended only as a toy model of how a phantom energy
f\/ield can emerge naturally from a gauge theory with a graded Lie algebra. A more experimentally testable
variation of the above toy model could have some coupling between
the scalar and vector f\/ields of the present model and the Standard Model f\/ields. Such a
coupling could be introduced in a phenomenological fashion via some {\it ad hoc} coupling. A more interesting
option would be to consider some larger graded algebra, such as $SU(N/1)$. Some of the f\/ields could be given
the standard assignment of even or odd generators (i.e.\ as in~\eqref{graded-alg}) while others could be given
the assignment in~\eqref{graded-alg2}. The f\/ields given the standard assignment would give standard gauge
f\/ields, while f\/ields given the non-standard assignment would give phantom energy and dark matter
f\/ields. This would give a new type of ``Grand Unif\/ied Theory'' with the phantom energy and
dark matter f\/ields replacing the extra gauge bosons of ordinary Grand Unif\/ied Theories.
Other authors \cite{wei} have used non-standard gauge groups such as $SO(1,1)$ to give models of phantom energy.

An important feature of the above model is the assumption that the Grassmann
vector f\/ields form permanently conf\/ined condensates.
This was a crucial to our phantom energy model since it
leads to a condensate of the $A^\pm _\mu$ and $B^\pm _\mu$ f\/ields. This in turn gave
a potential $V(\varphi ^8) =  \frac{v}{4}g^{2}(\varphi ^{8})^{2} $
for the $\varphi ^8$ f\/ield which was of the correct form to allow $\varphi ^8$ to act
as phantom energy. Aside from the present application to phantom energy one
might try to use the above mechanism to generate standard symmetry break by starting with
a graded Lie algebra but using all vector f\/ields rather than mixing vector and scalar. In
this way some of the vector f\/ields would be standard vector f\/ields, while other would be
Grassmann vector f\/ields. By the above mechanism the Grassmann vector f\/ields would form
condensates which would then give masses to the standard vector f\/ields i.e.\ one would have a
Higgs mechanism with only vector f\/ields).

An additional avenue for future investigation is to see if one could have a phantom energy model
with the original graded Lie algebra models (i.e.\ with vector f\/ields assigned to
even generators and scalars to odd) but using the supertrace. One would then have the problem
of some of the vector f\/ields having the wrong sign in the kinetic term, but this might
then give a~phantom energy model with a vector rather than scalar f\/ield.

As a f\/inal note the dark energy f\/ields (those connected with the even generators) considered here
violate all the known energy conditions which are normally required of quantum f\/ields. The reason
for taking this drastic step is that it gives a model for phantom dark energy which is thought to drive
the observed expansion of the Universe. If it turns out that the indications for $w < -1$ are
not correct (i.e.\ if $-1 < w < -1/3$) then there would be no need for phantom dark energy; ``ordinary''
dark energy would do. Recently, \cite{wilshire} there has been a proposal that dark energy ef\/fects
can be entirely explained by non-localized gravitational energy or rather gradients in gravitational
energy. If this proposal is correct then there would be no need that we can see for dark energy
in any form -- phantom or otherwise.

\subsection*{Acknowledgments} DS acknowledges
the CSU Fresno College of Science and Mathematics for a sabbatical leave during
the period when this work was completed, and a CSM 2007 Professional Development
Grant to attend {\it Symmetry--2007}.

\pdfbookmark[1]{References}{ref}
\LastPageEnding
\end{document}